\pdfoutput=1
\documentclass{JINST}
\usepackage{lineno}

\title{CHANTI: a Fast and Efficient Charged Particle Veto 
Detector for the NA62 Experiment at CERN}

\author{F. Ambrosino$^{a,b}$, T. Capussela$^{a,b}$, D. Di Filippo$^{a,b}$, P. Massarotti$^{a,b}$, M. Mirra$^{a,b }\footnote{Corresponding author.}$, M. Napolitano$^{a,b}$, V. Palladino$^{a}$, G. Saracino$^{a,b}$, L. Roscilli$^{b}$, A. Vanzanella$^{b}$, G. Corradi$^{c}$, D. Tagnani$^{c}$, U. Paglia$^{c}$\\
\llap{$^a$}Dipartimento di Fisica "Ettore Pancini" Universit\`a degli Studi di Napoli Federico II,\\
via Cinthia, Napoli, Italy\\
\llap{$^b$}INFN Sezione di Napoli, \\
via Cinthia, Napoli, Italy\\
\llap{$^c$}INFN Laboratori Nazioni di Frascati,\\
Via E. Fermi 40, Frascati, Italy\\
  E-mail: \email{Marco.Mirra@na.infn.it}}

\abstract{The design, construction and test of a charged particle detector made of 
scintillation counters read by Silicon Photomultipliers (SiPM) is described. The detector, which operates in vacuum and is used as a veto counter in the NA62 experiment at CERN, has a single channel time resolution of $1.14$ ns, a spatial resolution of $\sim2.5~$mm and an efficiency very close to $1$ for penetrating charged particles.}


\begin{document}

\section{Introduction}

\begin{figure}
\includegraphics[width=1.\textwidth]{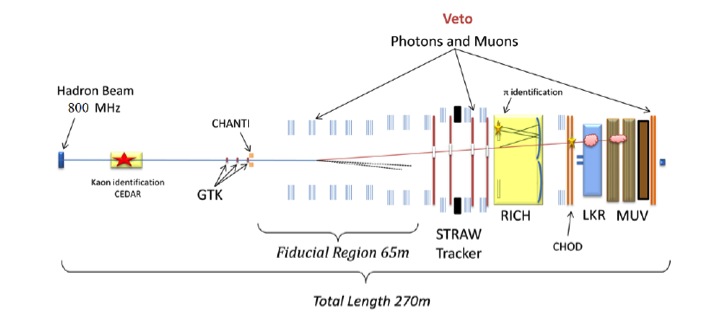}
\caption{Schematic layout of the NA62 apparatus.}
\label{na62}
\end{figure}

The study of flavour changing neutral current (FCNC) meson decays, forbidden
at tree level by the Standard Model (SM), allows one to test the SM and to search for
signals of new physics in a way complementary to the study of the processes
at very high energies, where the possible contribution of new particles is
expected to manifest itself already at "leading order". In this context, the
NA62 experiment at CERN~\cite{NA62} has the main purpose to measure with 10\% precision the branching fraction 
of the rare decay  K$^{+}\rightarrow
\pi ^{+}\nu \bar{\nu}$. The SM provides the fairly accurate
prediction BR$_{SM}($K$^{+}\rightarrow \pi ^{+}\nu \bar{\nu})
=( 9.11\pm 0.72) \times 10^{-11}$~\cite{MS}. At present the
experimental value is BR$($K$^{+}\rightarrow \pi ^{+}\nu \bar{\nu}~%
) =( 1.73_{-1.05}^{+1.15}) \times 10^{-11}$\cite{SP}; it is
based on the observation of seven events made by the experiments E797 and E949
at BNL~\cite{BNL}.
 NA62 plans to collect a sample of about $80$ SM events in two years of data
taking, studying the decay in flight of the  K$^{+}$ contained in an intense
secondary beam of positive particles of 75 GeV/c momentum, produced using
the proton beam of the CERN SPS. Since the two neutrinos cannot be observed,
the topology of the events is characterized by an incoming K$^{+}$ meson and
a positive track coming out from the decay vertex. Therefore, the
measurement strategy is based on the recognition of the incident kaon, the
measurement of its momentum and of the momentum of the outgoing
pion and on the rejection of $10^{10}$ times more abundant events due to
other kaon decay channels. The background will be rejected making use of
kinematic constraints, veto counters and detectors that allow particle
identification, in particular the $\pi /\mu $\ discrimination. The
individual detectors that make up the overall apparatus used
by NA62 are schematically represented in figure \ref{na62}.
A particularly insidious source of background is given by events produced by
the inelastic interactions of the beam with the three measurement stations -
in particular the third - of the beam spectrometer (GTK) ~\cite{TD}. The spectrometer,
which consists of thin silicon pixels detectors, is installed in vacuum immediately upstream 
of the decay vacuum tube in which is located the fiducial
region. A positive pion, produced by the beam impacting GTK  and detected by the
apparatus of NA62 so that its origin is erroneously reconstructed in the
fiducial region, mimics a signal event. To reduce this type of backgroud
to an acceptable level, a detector to veto charged particles was designed
and built (CHarged ANTI-Counter, CHANTI). It is made of scintillation
counters and is placed in vacuum immediately after GTK3 without
interfering with the beam. CHANTI exploits the fact that the events to be
rejected are normally inelastic events at high multiplicity of tracks. To
ensure a high efficiency of rejection it has just to be able to intercept
charged tracks produced at angles, with respect to the beam direction,
ranging from $49$ mrad up to $1.34$ rad; the region up to $49$ mrad is covered 
by the photons veto system.
The following describes the design and construction of CHANTI and the
results of tests carried out using both cosmic rays and a beam of positive
particles.

\section{Design and construction of CHANTI}

First of all CHANTI must be able to detect the inelastic interactions of the
beam with very high efficiency. Considering that it will be impacted not
only by secondary products of such interactions but also by the beam halo,
CHANTI has to be able to substain a high hit rate of charged particles, with a maximum rate per unit surface of the order of tens of
kHz/ cm$^{2}$ in the hottest region, and must have a radiation hardness up to
a few Gy/~year. A time resolution of $\sim 1$ns is also required in
order to have a "random veto" not exceeding few percent. Finally, since it will
be positioned in vacuum, the detector must have a small out-gassing rate and must
consume a relatively low power. In principle a specific tracking ability is
not required, however it is welcome in order to distinguish interactions due
to the halo from those due to the beam, both to monitor the halo itself in
the vicinity of the beam and to improve the time resolution without necessarily
increasing the granularity.

\subsection{Basic elements}
\begin{figure}[htbp]
  \centering
  \includegraphics[width=.28\textwidth]{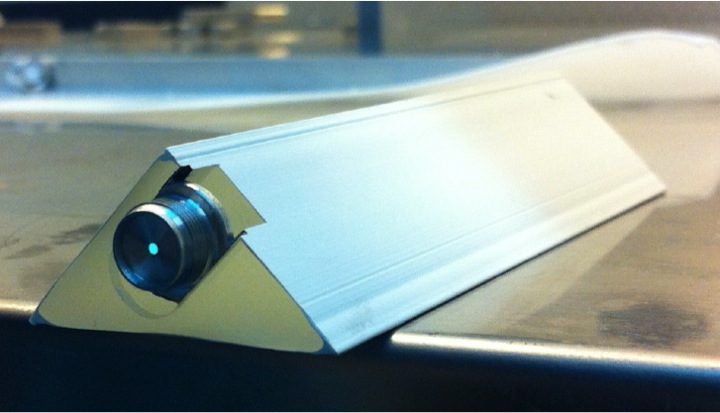}%
  \qquad
  \includegraphics[width=.32\textwidth]{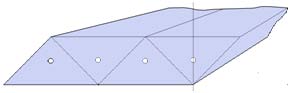}%
 \qquad
\includegraphics[width=.28\textwidth]{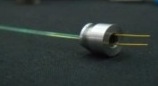}%
  \caption{Left: bar shape. Centre: bars arrangement. Right: Fiber connector with SiPM.}
\label{bar}
\end{figure}

With these requirements in mind, we have chosen to build the CHANTI using
plastic scintillator bars made of extruded polystyrene (Dow Styron 663 W)
doped with 1\% PPO and 0.03\% POPOP (by weight). It is an inexpensive
scintillator with emission in the blue, with a good yield of light, a rather
fast response ($\mathit{\tau} \sim $ few ns) and a good radiation resistance (5\%
degradation after a dose of $10^{4}$ Gy gamma irradiation). The bars,
produced by FNAL-NICADD~\cite{NICADD}, have a cross section in the shape of
an isosceles triangle with base 33 mm and height 17 mm (figure \ref {bar} on the left)
and carry a co-extruded TiO$_{2}$ coating $0.25$ mm thick. 
The bars, arranged next to each other in the way indicated in figure \ref{bar}, centre, 
allow to realize a plane counter 17 mm thick, essentially
with no fractures for tracks that pass through it. 

Each bar has longitudinally a central hole with a diameter of $1.7$ mm in
which an optical WLS fiber is housed. The fiber is the fast multiclad blue to green
shifter Saint-Gobain BCF-92 of 1 mm diameter. It has an emission spectrum
peak at $492$ nm, a decay time of $2.7$ ns and an attenuation  length of $3.5$
m. The free space between the fiber and the scintillator is filled with an
optical glue (SCIONIX$^{\circledR}$ Silicon Rubber Compound RTV615). It provides excellent optical
contact and has a very low outgassing rate~\cite{OUTGAS}.
The fiber is coupled at one end to a SiPM by means
of a connector (figure \ref{bar} on the right) bonded to the scintillator with a
high viscosity epoxy glue (Epoxy 3M DP490). The other end is made reflective by a
sputtered Al film. The connector is designed and constructed so as to ensure
an optimal and stable coupling between the fiber and the SiPM. It is made of two
parts: a base, in which the SiPM is housed, glued both to the scintillator
and to the fiber, and a screw cap by which the SiPM is blocked so as to
keep the optical coupling with the fiber itself. Both parts are made of
aluminum and machined to a high precision.
The SiPM is a ceramic package Hamamatsu S10362-13-050C, which has a sensitive area of 
1.3$~\times$ 1.3~mm$^{2}$. It is a photodetector of the type MPPC (Multi-Pixel
Photon Counter) consisting of a matrix of Avalanche Photo-Diode (APD) that operate in Geiger regime.
The single cell has dimensions $50\times 50~ \mu$m$^{2}$; the total number
of cells is 667. The specifications provided by the manufacturer and
referred to a temperature of operation of 25$^\circ$C are: 
\begin{itemize}
\item  operation voltage typically $70$ V;
\item photon detection efficiency (PDE), as quoted by manufacturer, slightly less than $50$\% for photons of $440$ nm - peak of the spectral response - and higher than $40$\% between 370 and 520 nm;
\item gain $7.5$~$\times$ 10$^{5} $;
\item dark counting rate of the order of $800$ kHz at a threshold of 0.5 photoelectron\footnote{For simplicity we use here and in the following the term
photoelectron to indicate the primary electron (hole) generated by a
photon hitting the SiPM.} (pe).
\end{itemize}

The SiPM was choosen after testing samples of three different photodetectors from the
Hamamatsu S10362 series, namely: 11-050C, 11-100C and 13-050C. The first
two differ from 13-050C as they have a sensitive area of $1\times 1$ mm$^{2}$
with a cell $50\times 50~\mu$m$^{2}$ the first and $100\times 100~\mu$m$^{2}$
the second. We measured the responses of various devices coupled to the
same bar of scintillator exposed to a collimated source of $^{90}$Sr. The
measurements were carried out in a thermostatic chamber maintaining a fixed
temperature of 25${^\circ}$C with an uncertainty better than 0.1${^\circ}$C .

The SiPM signals, amplified x10, were fed into an oscilloscope
Tektronix TDS5054 read out by using a GPIB connection and a LabView program. The 500 MHz
bandwidth of the oscilloscope is more than sufficient to analyze signals
with rise times of the order of a few ns. In order to select the SiPM we used as a
figure of merit the signal amplitude normalized to the signal of single
photoelectron, that was determined in advance by collecting dark noise
signals~\cite{TD}. The samples of the series 13-050C gave a number of
photoelectrons between 20 and 25,  systematically greater than that given by
the SiPM from the other two series.
SiPM, like all semiconductors, may
be damaged when exposed to an intense neutron flux, however they remain
substantially unaltered after irradiations not greater than about 2$\div
$3$~\times$ 10$^{9}~neq~/$cm$^{2}$ ($neq=$ 1MeV neutron equivalent) \cite{RadR}. We
checked by two different simulations, one made by FLUKA~\cite{FLUKA} and
the other using GEANT4~\cite{GEANT4}, that this limit will not be exceeded
in the two years of data taking foreseen for NA62. Indeed, the dose of
neutrons integrated by the CHANTI SiPM is expected to be less than $4\times
10^{8}~neq~/$cm$^{2}/ y $~\cite{TD}.

\subsection{Layout and assembly}
\begin{figure}
\includegraphics[width=1.\textwidth]{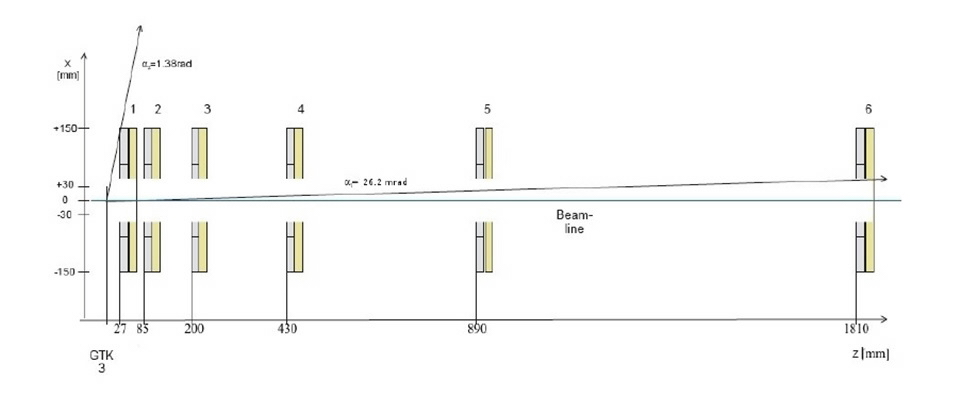}
\caption{CHANTI stations layout with respect to GTK3 position.}
\label{layout}
\end{figure}

CHANTI is made of six square counter hodoscopes (stations) of 30 cm side
arranged in vacuum along the beam direction at distances of 27, 85, 200,
430, 890 and 1810 mm from GTK3 (figure  \ref{layout}  ). 
In order not to hinder the beam, the stations have a central hole of 
65$~\times$ 95 mm$^{2}$ rectangular shape, with the longer dimension in the
horizontal direction (x-axis) and the shorter in the vertical direction
(y-axis). The geometry of the system is such that the CHANTI can intercept
secondary particles, generated by beam interactions at the center of GTK3,
which propagate in an angular region from 26.2 mrad to 1.38 rad with respect
to the beam axis and has full acceptance for those running between $49$ mrad and $1.34$ rad.

\begin{figure}
\centering
\includegraphics[width=.5\textwidth]{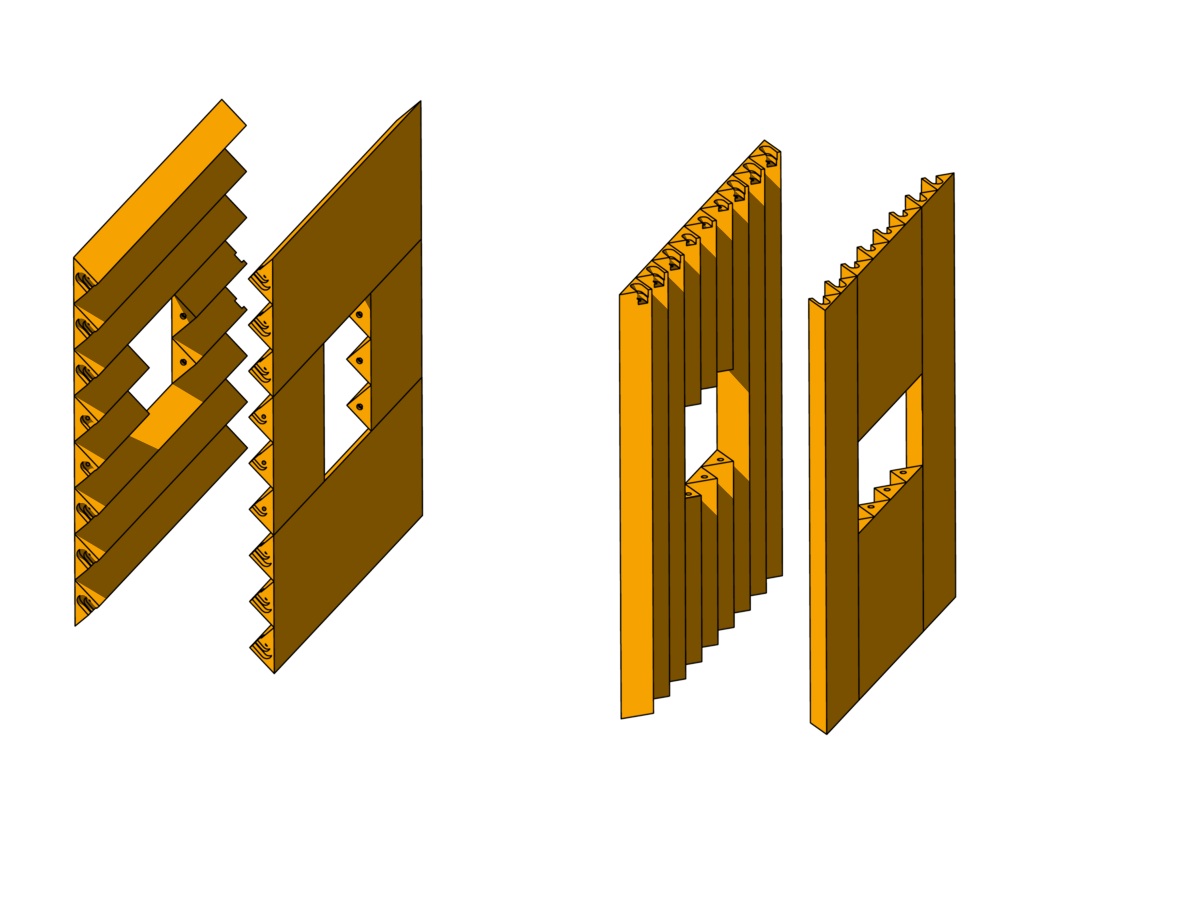}%
\includegraphics[width=.5\textwidth]{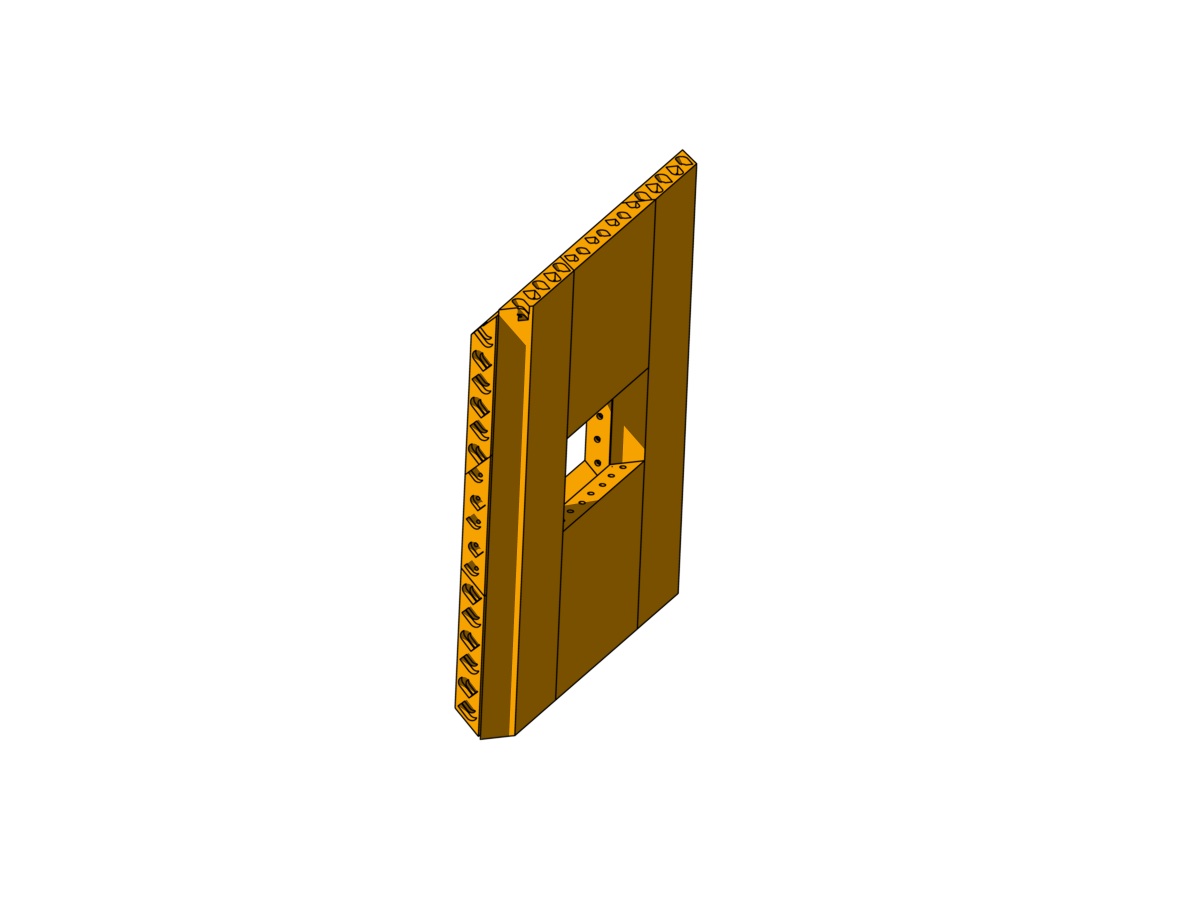}
\caption{Left: Layer's structure. Right: CHANTI station.}
\label{station}
\end{figure}

Each station contains 46 bars of variable length, divided into two separate
flat layers of scintillator, in contact with each other, formed by coupling
and glueing the triangular bars in the manner specified earlier and shown in
figure  \ref{bar} in the centre. One of the layers has the bars arranged in the
horizontal direction (Y layer, 22 bars) and the other has the bars in the
vertical direction (X layer, 24 bars). Because of the central hole we used bars of different lengths: long bars (300 mm), medium bars
(117.5 mm) and short bars (102.5 mm). The X layer is composed of 10 long and
14 medium bars while the Y layer of 12 long and 10 short bars (figure \ref{station}).
The hit rate that each bar must support varies according to the position of
the bar and along the bar itself. This rate has been carefully studied by
means of a MonteCarlo (MC) simulation, as a high rate induces an inefficiency in data
collection because of the SiPM recovery time. The two horizontal long bars
(Y layer) closer to the hole for passage of the beam
substain the highest rate. In order to keep the maximum rate of each SiPM in
the limit of $\sim $1 MHz, each of these bars has been divided into two
halves, each one read by a separate SiPM. This causes the number of SiPM and,
therefore, the number of readout channels of each station, to become 48. To minimize dead zones, the bars were divided with a cut transverse to their
longitudinal axis and the two facing ends were painted with titanium dioxide
in order not to have cross-talk.
The realization of each station was done following three main steps: 
\begin{enumerate}
\item  preparation of the bars with their fibers and connectors for SiPM, 
\item quality test of the bars,
\item assembly of the station using only bars that passed the test.
\end{enumerate}
The first step starts bonding the SiPM connector to each
fiber, previously cut to the right length and mirrored by Al sputtering at the end opposite to SiPM.
Once ready, the fiber is inserted into the hole of the bar, the connector is glued to the bar and the hole
containing the fiber is filled with the optical glue, that was prepared in
advance in the amount appropriate to the lenght of the bar. During the
filling the bar was kept vertical and the glue was injected from the
bottom. Both the preparation and the injection of the glue were carried out
with great care to avoid the formation of air bubbles.
Once completed with fibers and connectors, the bars were characterized by
a quality test and accepted or not on the basis of the obtained results.
This characterization, which will be described in section \ref {bar_test}, is
important because a bar that was used for the realization of a station can
no longer be replaced should it be faulty. In fact, since all elements
which constitute a station are glued to each other, any apparatus malfunction
can be remedied only by replacing the entire station.
\begin{figure}
\centering
\includegraphics[width=.6\textwidth]{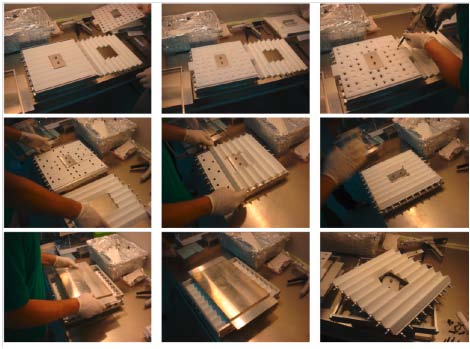}
\caption{Steps of assembly. From left to right and top to bottom:  half layer is arranged on the jig; teflon mask is aligned; glue spots are applied; mask is removed; bars of the other half layer are placed;  second jig is applied on top to align last half layer and to distribute pressure; half module as it appears the day after gluing.}
\label{assembling}
\end{figure}
When all bars were ready, a station was assembled in two distinct stages,
each of which took about one day. During the first day the central half-X  and half-Y layer
were glued together in the following way. First of all the bars of
the first half-layer were arranged, with the vertex at the bottom, on a
special aluminum jig that guaranteed the correct positioning (figure \ref{assembling}). A
series of glue droplets, each of about 0.1 ml of glue, were
distributed using a mask on the plane formed by the base surfaces of the
bars. Soon after that the bars of the second half-layer were positioned on that
plane using a second template. The two half-layers were then loaded with a
weight in order to exert an appropriate pressure to ensure a uniform
distribution of the glue and a good bonding of the parts. The second day the complementary bars were glued with a similar procedure, thus completing the assembly. A total
of about 20 ml of low out-gassing epoxy glue (3M DP490) was used for each
station. 

\subsection{Test of the bars \label{bar_test}}
\begin{figure}
\centering
\includegraphics[width=.75\textwidth]{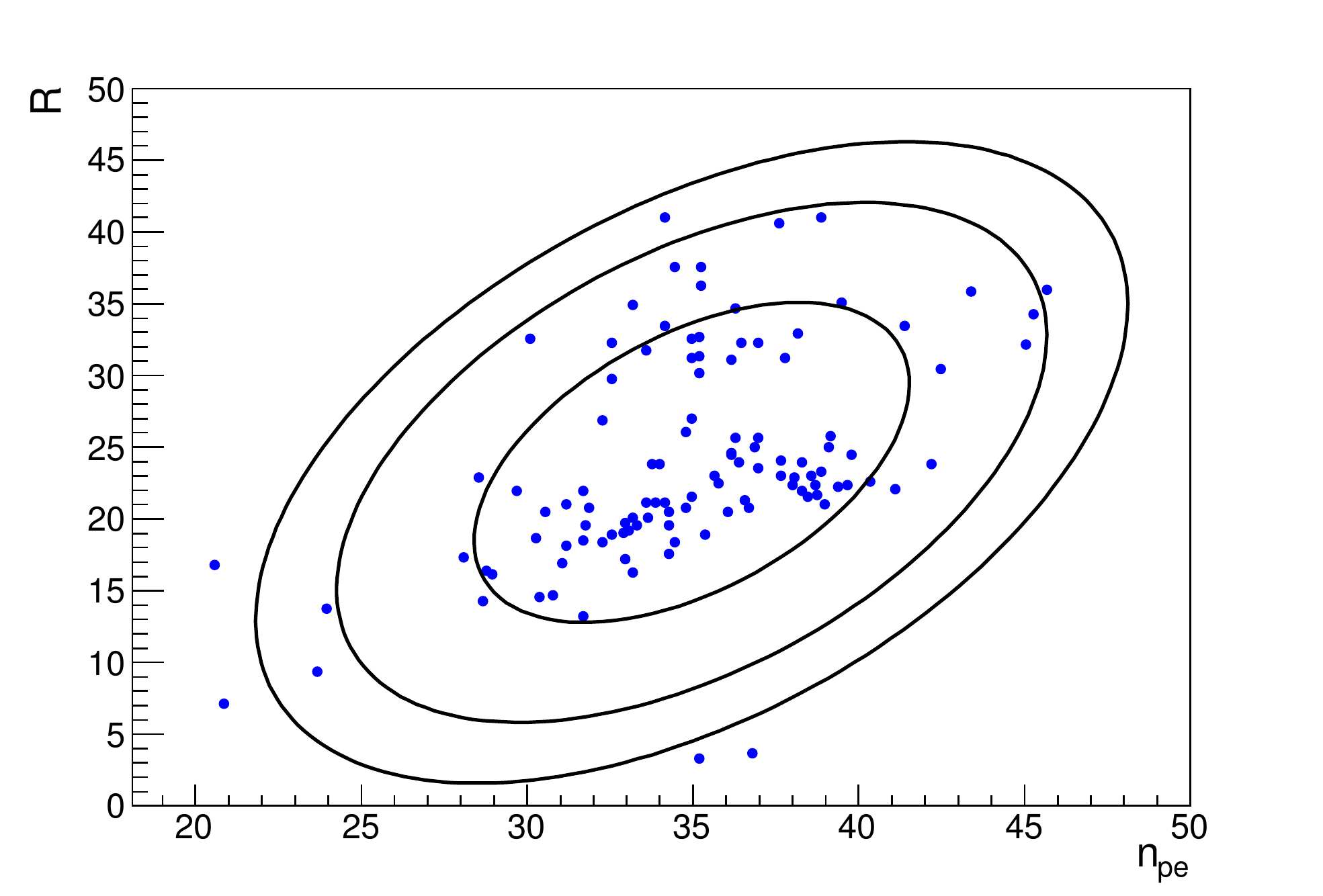}
\caption{The result of tests of  long bars.}
\label{plot_test_barre}
\end{figure}

The quality test of the bars and the characterization of each bar were made by
coupling the bar to a SiPM and studying its response to cosmic rays. The
test was performed in two distinct ways: the first in self-trigger mode and
the second by means of a trigger obtained using a telescope of two small
scintillation counters (surface 2.5$\times$ 2.5 cm$^{2}$) in coincidence. In
both cases, the bar and the readout electronics were placed in a
thermostatic chamber that allowed to maintain a constant temperature of 25
$\pm$ 0.1 $^\circ$C.
With the self-trigger test a comprehensive study was done of the response of
the bar to cosmics going through in every part of the bar and in any
direction. This allowed to have a sufficiently high rate of events, thus
reducing considerably the measurement time. The signal of the SiPM, 
amplified by a trans-resistance amplifier, was collected by means of a
fast digital oscilloscope Tektronix$^{\circledR}$ TDS5054 connected to a PC. 
In order to get a rate of acquisition dominated by
signals produced by cosmic rays, each signal was stored only if its
amplitude exceeded 80 mV (corresponding to 16 photoelectrons), so that the contribution from the thermal noise of
SiPM was negligible.  A fixed number of signals was
acquired for each bar and the bar was characterized by calculating the ratio
R ($\leq$ 1) between the number of signals exceeding the threshold of
250 mV (corresponding to 50 photoelectrons) and the total. For the second type of test the trigger is given by the telescope of counters in order to select cosmic rays (muons) traveling approximately in the vertical direction. It was placed above the bar at the opposite side
with respect to the SiPM. This way the bar responds to minimum ionizing particles
crossing nearly perpendicularly to its longitudinal axis. To reduce the time required to
test them all, the bars were tested two at a time, arranging one above the
other below the trigger telescope. Again, the signals produced by the two
SiPMs were amplified, digitized by
the oscilloscope and acquired by a PC (as before), if they exceeded
the threshold of 200 mV. Each signal was integrated with respect to time and
a number of photoelectrons ($n_{pe}$) produced by the event was calculated by
comparison to the area of the signal of a single photoelectron. The
latter had been previously evaluated by studying the thermal
noise produced by the SiPM not coupled to the bar. The bar was characterized
using the average number of photoelectrons $n_{pe}$. After completing the
measures, a two-dimensional plot (figure \ref{plot_test_barre}), 
in which each bar is represeted by a point at coordinates $(
n_{pe}$,~R), was built for each type of bar. 
Three ellipses were drawn on the plot, corresponding to variations of 
1$\mathit{\sigma} $, 2$\mathit{\sigma} $ and 3$\mathit{\sigma} $ with respect to the average values 
($\bar n_{pe}$,$~\bar R$), obtained after a rotation done to find the two uncorrelated
variables. About $6$\% of produced bars that fell outside the 2$\mathit{\sigma} $ contour and gave
$n_{pe} < \bar n_{pe}$ and R $< \bar R$ were discarded.

 \section{Read-out electronics} 
\begin{figure}
\centering
\includegraphics[width=.8\textwidth]{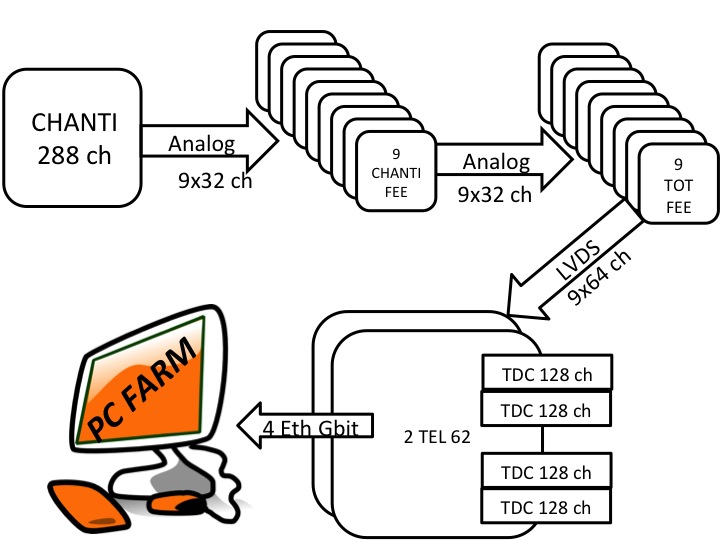}
\caption{Block diagram of the CHANTI read out elecronics.}
\label{schema_elettronica}
\end{figure}

Due to the triangular shape of the bar cross-section, even particles
crossing a bar with a trajectory perpendicular to the plane of the stations
pass through variable thicknesses of scintillator; so the signals
produced by SiPMs have, after amplification, amplitudes that typically
vary between few mV to hundreds mV, with rise times of $\sim 6 $ns. The
inner bars, being closer to the beam, give the highest rate of signals
entering the front-end electronics. MC simulations showed that the expected
value for that rate is of the order of 1 MHz.
 A block diagram of the read-out electronics, which is based on three
different types of 9U VME boards, is shown in figure \ref{schema_elettronica}. 
The first board has three different functions:
\begin{enumerate}
\item  provides for the SiPM V-bias and monitoring of SiPM current,
\item checks the temperature stability of SiPM reading Pt100 probes (platinum resistance thermometers), 
\item receives signals from SiPM and transfers them to the next stage through trans-resistance amplifiers.
\end{enumerate}
Each card has 32 channels that enter and leave through sub-DB (37 pin) connectors, and is able to monitor 4 two-pin Pt100 probes.
In order to achieve the necessary stability of the SiPM performance, each channel must have a stability in the bias voltage of order 10 mV or better for typical bias voltages of order 70 V. The dark currents, which at room temperature and in normal operating conditions are typically of order 100- 200 nA, are read with an accuracy of 1 nA. Analog to digital conversion to read the channel voltage and currents is made via AD7708 16 bit ADC for each channel, while the applied voltage is controlled and set individually for each SiPM via LTC2620 12 bit DAC.
The reading of the Pt100 probes is used in the software control system of the detector to adjust the bias voltage of each SiPM in order to mantain it at a fixed working point. The fast amplifier, providing a factor 25 on signal amplitudes for 50 Ohm impedance, has a cutoff frequency of 80 MHz. It is chosen to reach a reasonable compromise between the need to follow the risetime of the signals and the necessity to keep to an acceptable level high radio-frequency noise.\\
The second board ("$ToT$ board") is adapted from LAV (Large Angle Vetoes)
read-out electronics of NA62 \cite {LAVelectr}. Each input analog signal is filtered by means of an amplifier/splitter that produces two identical copies of the signal and feeds them to two discriminators. Each discriminator converts the analog input into a LVDS output whose duration equals the time the input signal stays above a configurable threshold. Therefore each input analog signal corresponds to two LVDS output signals. Since two different values of the  thresholds are used, these two LVDS output signals have different durations. This arrangement is useful for correcting for time-walk (see section \ref{detector_performance}). The 64 LVDS signals produced this way are sent to TDCs contained in the third board
using two SCS12 connectors placed on the front of the card. The analog input
signals are also added four by four and sixteen by sixteen; the result is
available on 8 + 2 connectors LEMO00 for the purpose of monitoring. The
setting of the thresholds and the communications with DAQ card are
operated with a CANopen protocol through two RJ-45 connectors.

The third board (TEL62, Trigger Electronics for NA62), common to most of
NA62 detectors, is a general purpose trigger and data-acquisition board that
manages the read-out of NA62 \cite {TEL62}. It receives the LVDS logic
signals by custom designed TDC mezzanines that provide  the
leading and trailing times for each signal with an accuracy of 100 ps LSB. The motherboard
houses 4 TDC mezzanines, each serving 128 input channels. Each of the four
TDC mezzanine is served by a FPGA (PP-FPGA); a fifth FPGA (SL-FPGA)
collects data from the first four and drives another mezzanine board housing
a 4-Gigabit Ethernet link. The TEL62 provides both for data acquisition and
for the generation of primitives\footnote{Trigger primitives are data items
from which the L0 trigger takes the L0 trigger decision.} from most of the detectors used for the L0
trigger and sends the processed data to the PC farm in response
to the L0 trigger. The TEL62 is based on the design of the TELL1 read-out card 
\cite {TEL1}, developed for the LHCb experiment, of which essentially retains
the basic functions but with modifications to suit the needs of NA62 and to
take into account developments of electronic components post TELL1 project.

\section{Expected detector performance}\label{detector_performance}
\begin{figure}[htbp!]
\centering
\includegraphics[width=.6\textwidth]{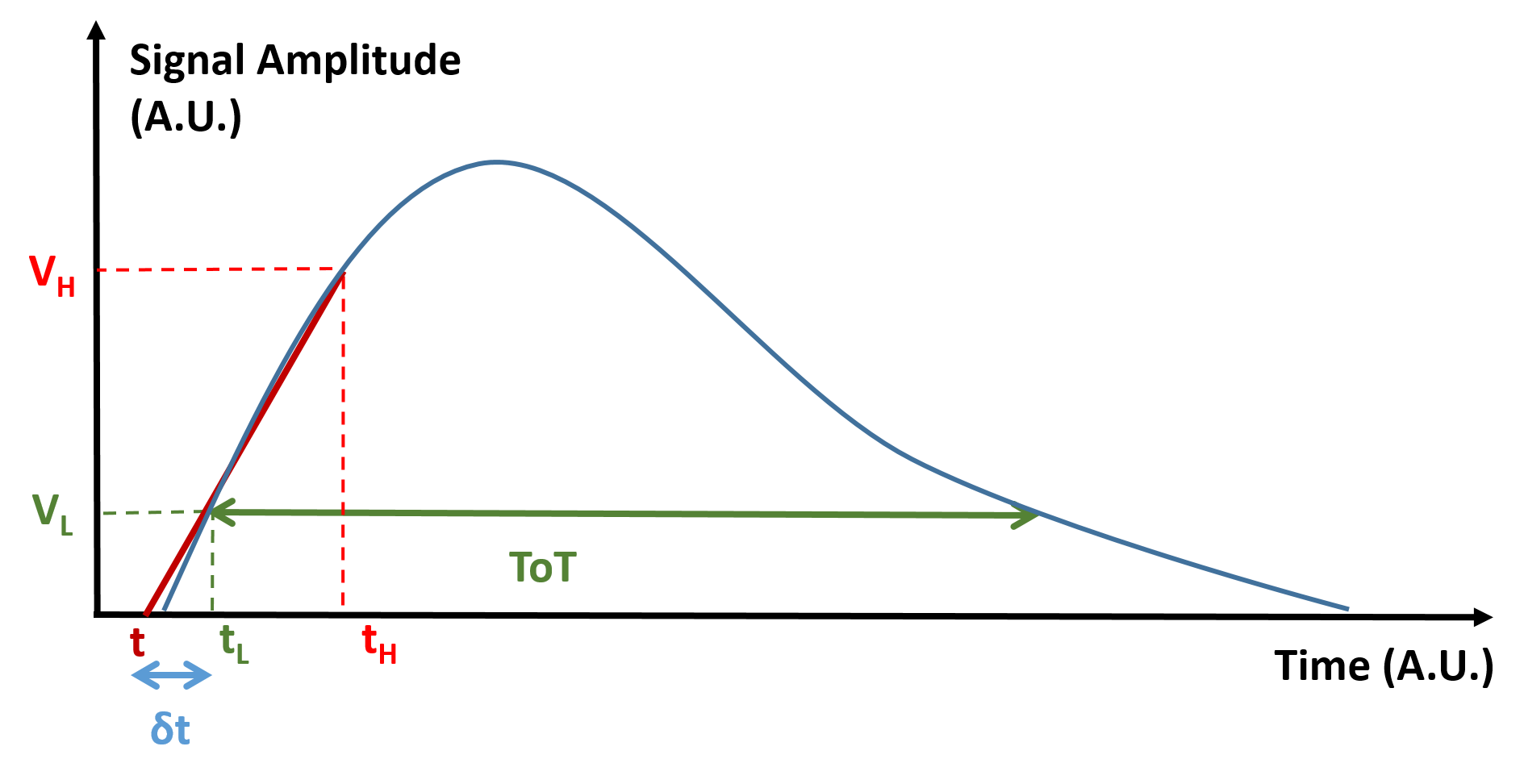}
\caption{Time-walk correction $(\delta t)$ in linear approximation (eq.~\protect\ref{TimeWalkCorr}).}
\label{SignalShape}
\end{figure}

\begin{figure}
\centering
\includegraphics[width=.6\textwidth]{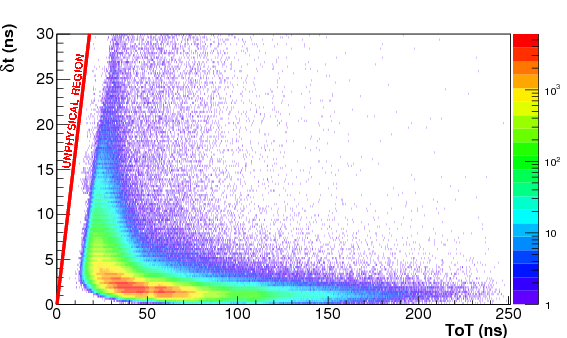}
\caption{Time walk correction vs $ToT$ for events passing both thresholds.The unphysical region, at the left of the red line,  is defined by the condition t$_{H}$ - t$_{L}$ > $ToT$. }
\label{corr_t_walk_tot}
\end{figure}

\begin{figure}
\centering
\includegraphics[width=.6\textwidth]{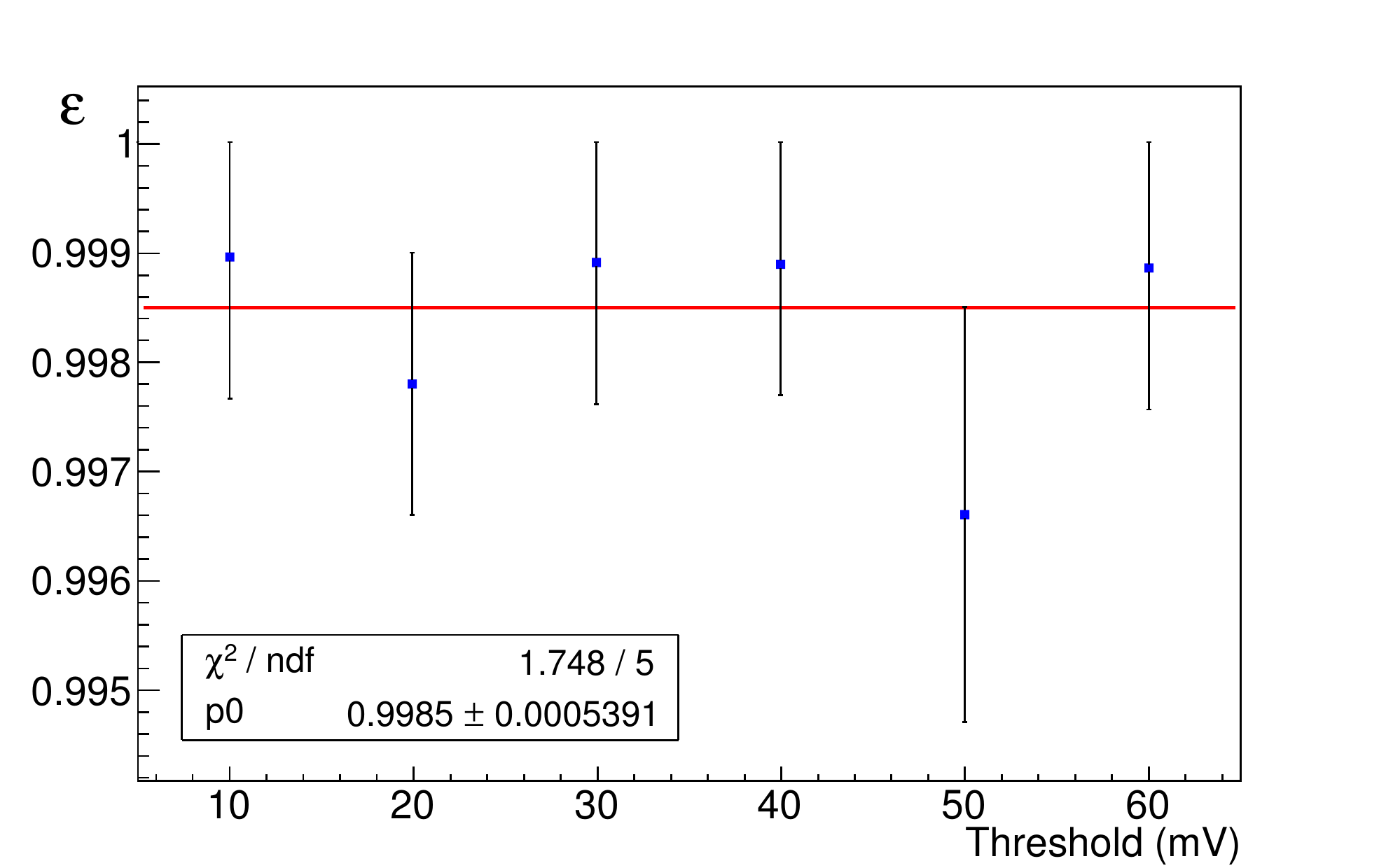}
\caption{Layer efficiency, measured with a prototype made of 5 bars, as a function of threshold.}
\label{eff_mimmo}
\end{figure}

The performance of the detector has been thoroughly tested on several prototypes, both using cosmic rays and during the so-called NA62 ``technical run'' in 2012. Tests have concentrated on the main figures of merit for the detector, namely its time resolution and its efficiency.  
Regarding the time resolution, the basic idea to reach the needed performance, is the ability to perform the appropriate time walk correction, without measuring the integrated signal charge. This is done using two features of the front-end electronics, namely the possibility to measure the time at which the signal crosses two different thresholds (so called Low and High threshold respectively), and the time it stays above each of them.
Signals passing both the low and high thresholds allow the determination of
the time t a particle passes through the detector corrected for the
"time-walk"; in fact, in linear approximation,
\begin{equation}
t=t_{L}-\delta t=t_{L}-V_{L}\frac{t_{H}-t_{L}}{V_{H}-V_{L}}
\label{TimeWalkCorr}
\end{equation}
where $V_{H}$ and $V_{L}$ are the high and low thresholds and $t_{H}$ and $%
t_{L}$ are the "leading times" corresponding to the high and the low
thresholds (see figure \ref{SignalShape}).\\
The $\delta t$ correction is correlated to $ToT$ as can be
seen from the distribution in figure \ref{corr_t_walk_tot}. In order to apply the time-walk correction also to the signals which only cross the low threshold, the distribution in figure \ref{corr_t_walk_tot} has been fitted and the $\delta t$ correction has been parametrized as a logarithmic function of the $ToT$.
Details of the tests performed have been reported elsewhere ~\cite{GIULIO},~\cite{MIMMO},~\cite{TechRun}. It is sufficient here to stress that tests done in different conditions, using beams or cosmic rays, have all pointed towards a single
 hit time resolution of the CHANTI (after time walk and position corrections are applied) of order 1 ns (cfr. $1.03 \pm 0.01$ in ~\cite{TechRun}). 
Regarding the efficiency, the first laboratory measurement was carried out using a prototype made of 5
bars. A 4-fold coincidence of small scintillator counters selected vertical muons crossing the three central bars of the prototype. The signals from the five bars were processed by the CHANTI front-end card and
sent to the "$ToT$" card which converted the analog input
signals into LVDS digital signals. The LVDS signals were then processed by a
TDC VME module (equipped with the same TDC of the TEL62 card) whose outputs
(leading and trailing times) were acquired by a PC. A good event was defined by at least
one bar fired giving an analog signal over a fixed threshold and a corresponding full LVDS signal, i.e.
a signal with a leading and trailing times, in the time window
of the trigger. Data were collected for 6 monospaced different thresholds ranging from 10 to 60 mV.
Events were collected at each threshold.
Figure  \ref{eff_mimmo} shows the efficiency as a function of the threshold. 
As can be seen the efficiency is the same, within error, for all thresholds and its mean value is $0.9986 \pm 0.0005$. Anyway the two lower thresholds, which correspond to 
a signal of about $1 \div 2$ pe, are not usable in practice in the final experimental setup; indeed, given the SiPM thermal noise, they would generate a single channel accidental activity in the CHANTI of the order of (or even above) the physical signal rate at full intensity. 

\section{Results from the first NA62 run}
\begin{figure}[!]
\centering
\includegraphics[width=.6\textwidth]{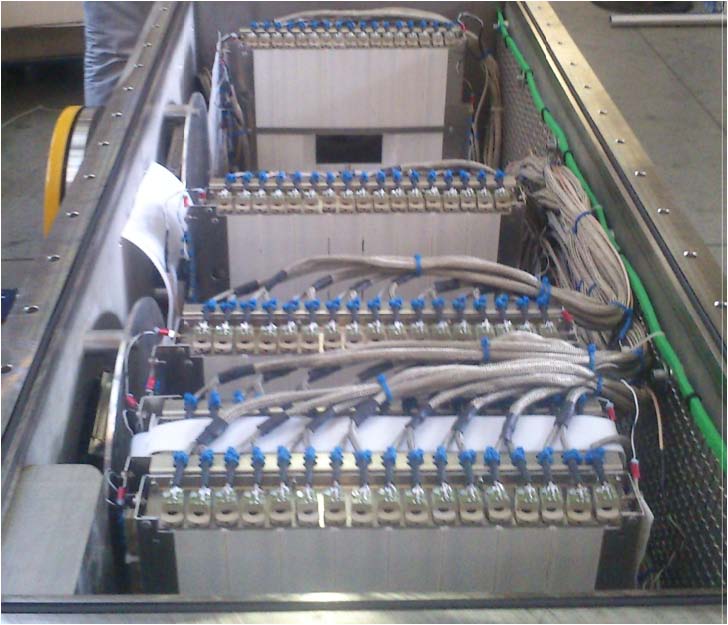}
\caption{The first five CHANTI stations assembled in their vacuum vessel.}
\label{photo_CHANTI}
\end{figure}

In the course of 2014 the construction of CHANTI was completed. Figure  \ref{photo_CHANTI}
shows the first five CHANTI stations assembled in their vacuum vessel. 
The sixth stays in a vacuum chamber extention at about one meter from the fifth. 
The whole detector, consisting of six separate stations (figure  \ref{layout})
for a total of 288 physical channels and 576 electronic channels, was
installed at CERN on the NA62 beam line. The experiment started data taking
with a run carried out between October and December 2014. The main purpose
of the run was the "commissioning" of the NA62 apparata. Data were collected
in three different conditions:
\begin{itemize}
\item   in the absence of beam,
\item  with a beam of hadrons,
\item with a beam of muons.
\end{itemize}

The third condition corresponds to the situation of beam but with the
presence of a "filter" according to which the particles that reach
the detector are essentially muons.

\begin{figure}
\centering
\includegraphics[width=.6\textwidth]{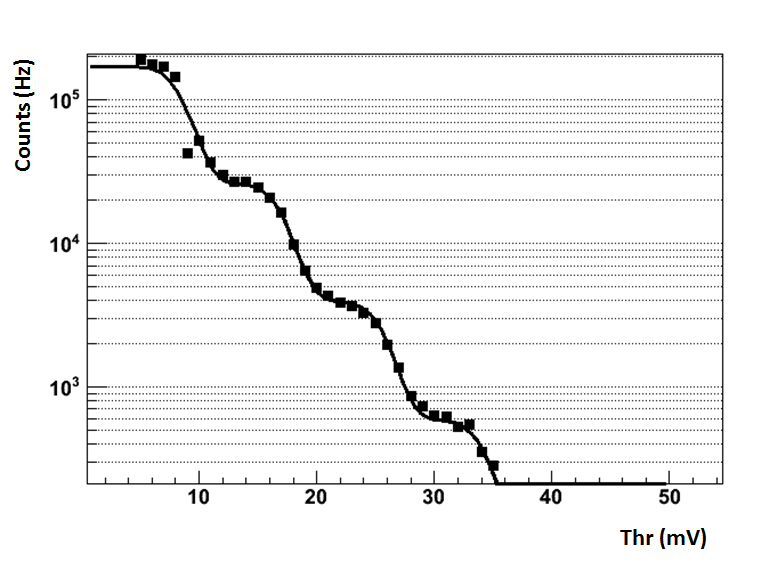}
\caption{A typical threshold scan curve. The distance between two consecutive inflection points measures the single photoelectron amplitude. For this particular channel it is found to be about 10 mV.}
\label{amp_dark_noise}
\end{figure}

\begin{figure}
\centering
\includegraphics[width=.6\textwidth]{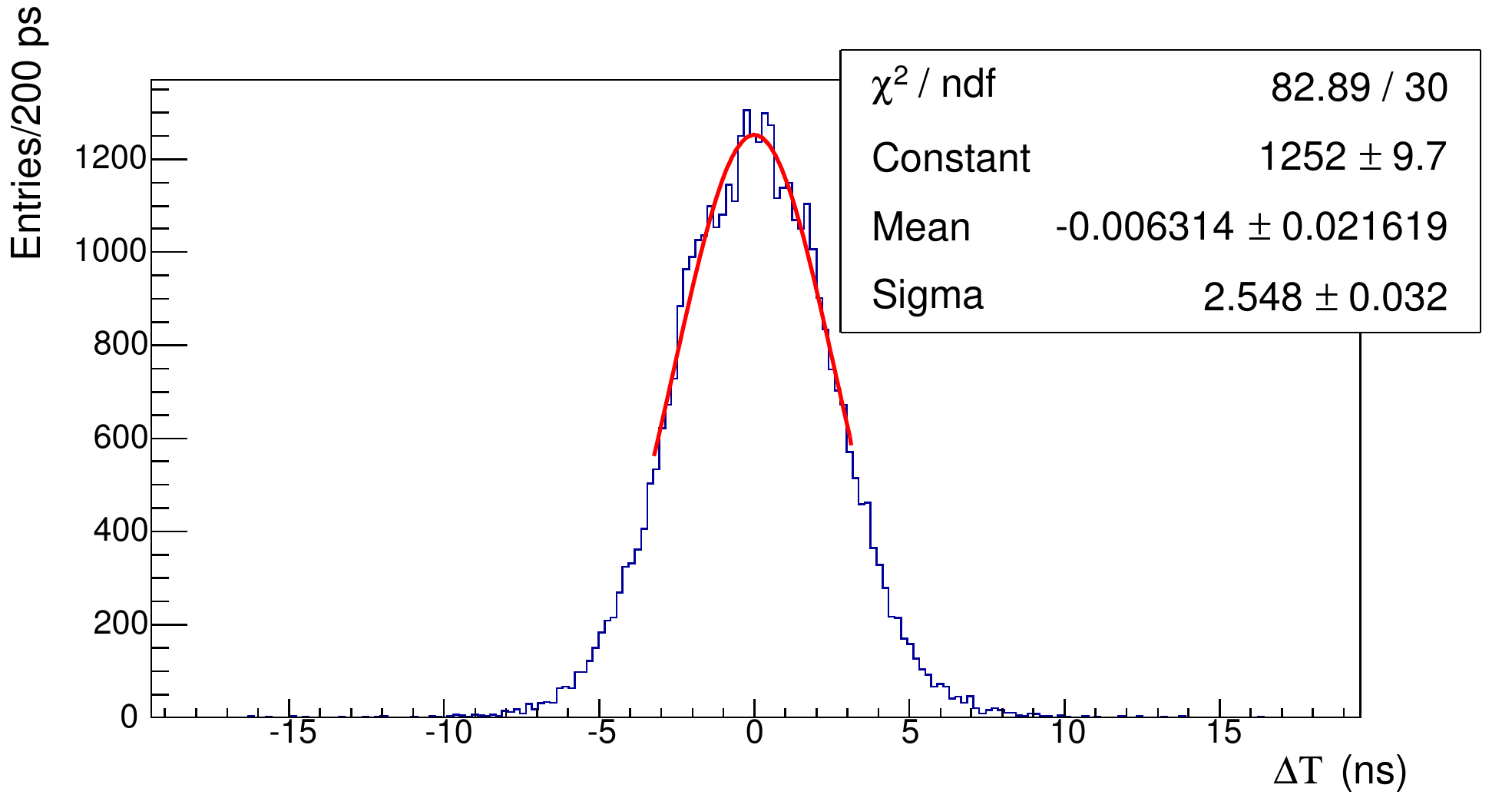}
\caption{Difference between the "leading times" of signals from two adjacent bars.}
\label{run_diff_leading_t}
\end{figure}

Data on dark noise (figure \ref {amp_dark_noise}), collected from the six
stations in the absence of beam, allowed to evaluate the amplitude
of the single photo-electron signal with the complete electronic chain.
This amplitude is equal to $\sim 9$mV for all channels in full
agreement with what was obtained previously.
Using part of the collected data, the time resolution of the detector was
re-measured by the method of the difference between the "leading times" of
signals from bars produced by the same track. A plot of this difference, not corrected for
time-walk and $t_{0}$ difference of channels, is shown in figure \ref {run_diff_leading_t}. 

\begin{figure}
\centering
\includegraphics[width=.6\textwidth]{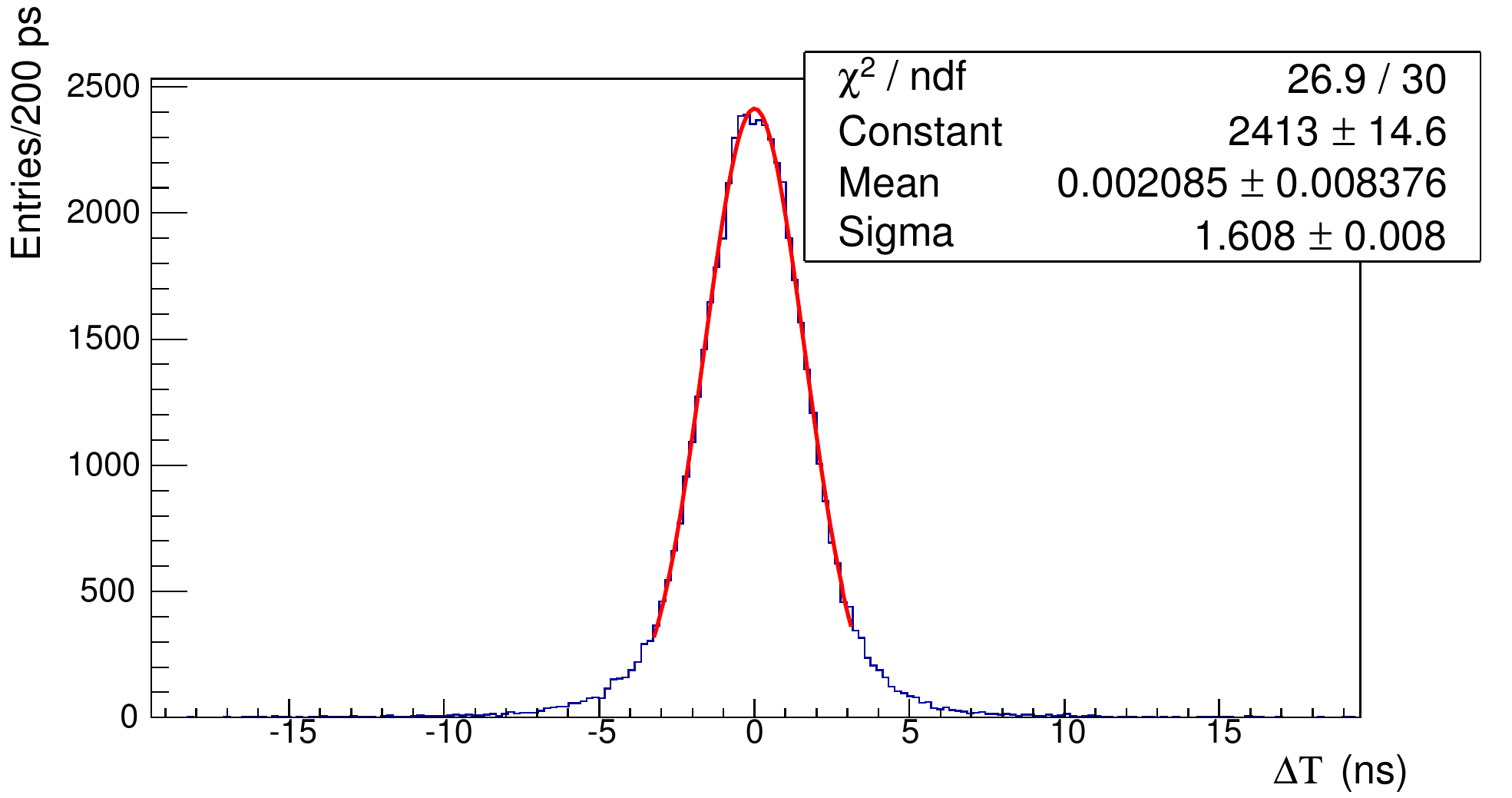}
\caption{Difference between the "leading times" of signals from two adjacent bars after time walk and reference detector time corrections.}
\label{run_corr_diff_leading_t}
\end{figure}

The first correction was parameterized according to the $ToT$ with the method
illustrated before; the second was estimated by the difference in the
transit time of particles of the same event recorded by CHANTI and by the
Cherenkov kaon tagger of NA62 (KTAG). Figure \ref {run_corr_diff_leading_t}
shows the corrected distribution, from which one gets a time resolution of $1.14$ ns, 
in agreement with the result of tests on prototypes.

\begin{figure}
\centering
\includegraphics[width=.6\textwidth]{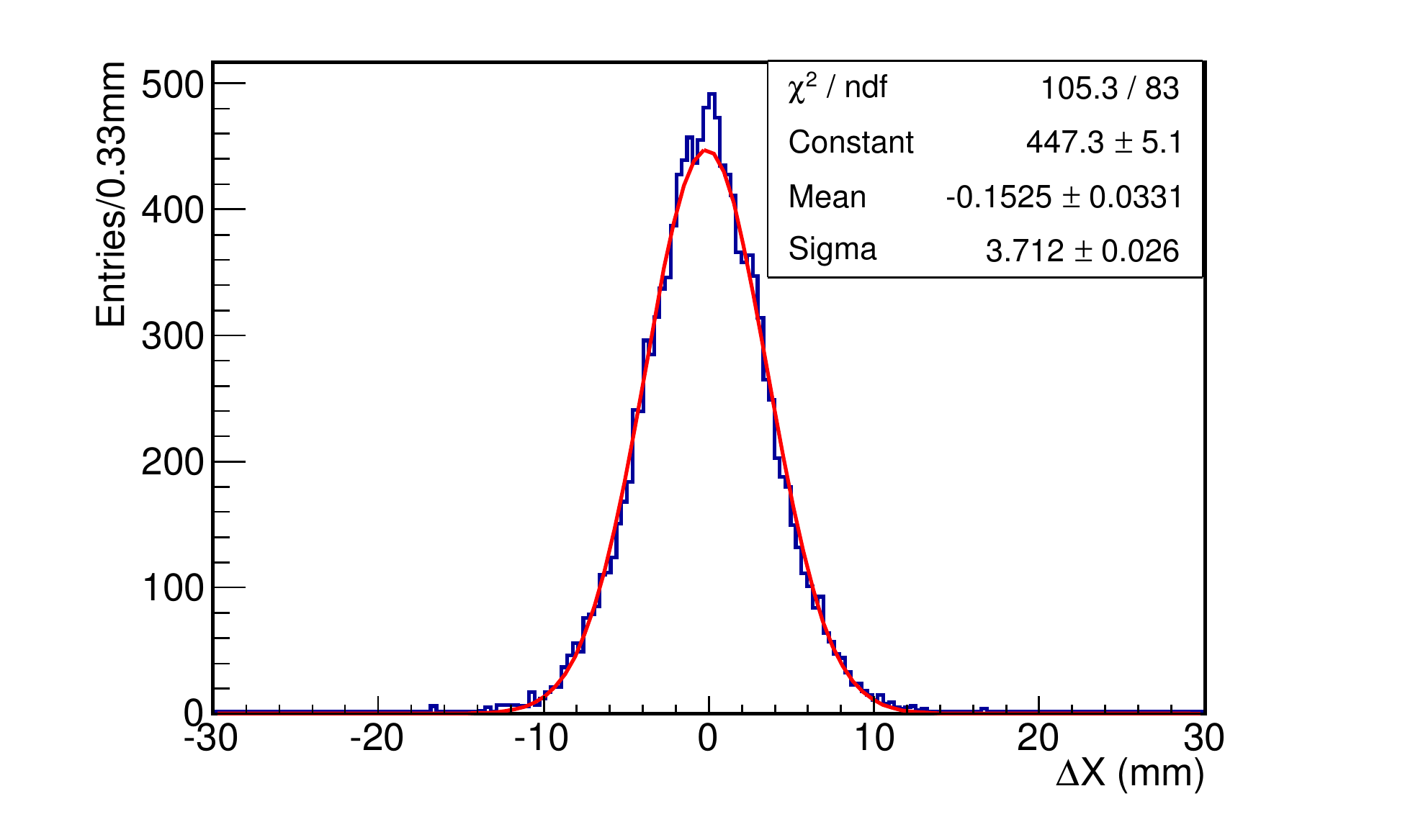}
\caption{Difference between X position of two hit points on different station for tracks parallel to beam direction.}
\label{diff_2_5}
\end{figure}

\begin{figure}
\centering
\includegraphics[width=.45\textwidth]{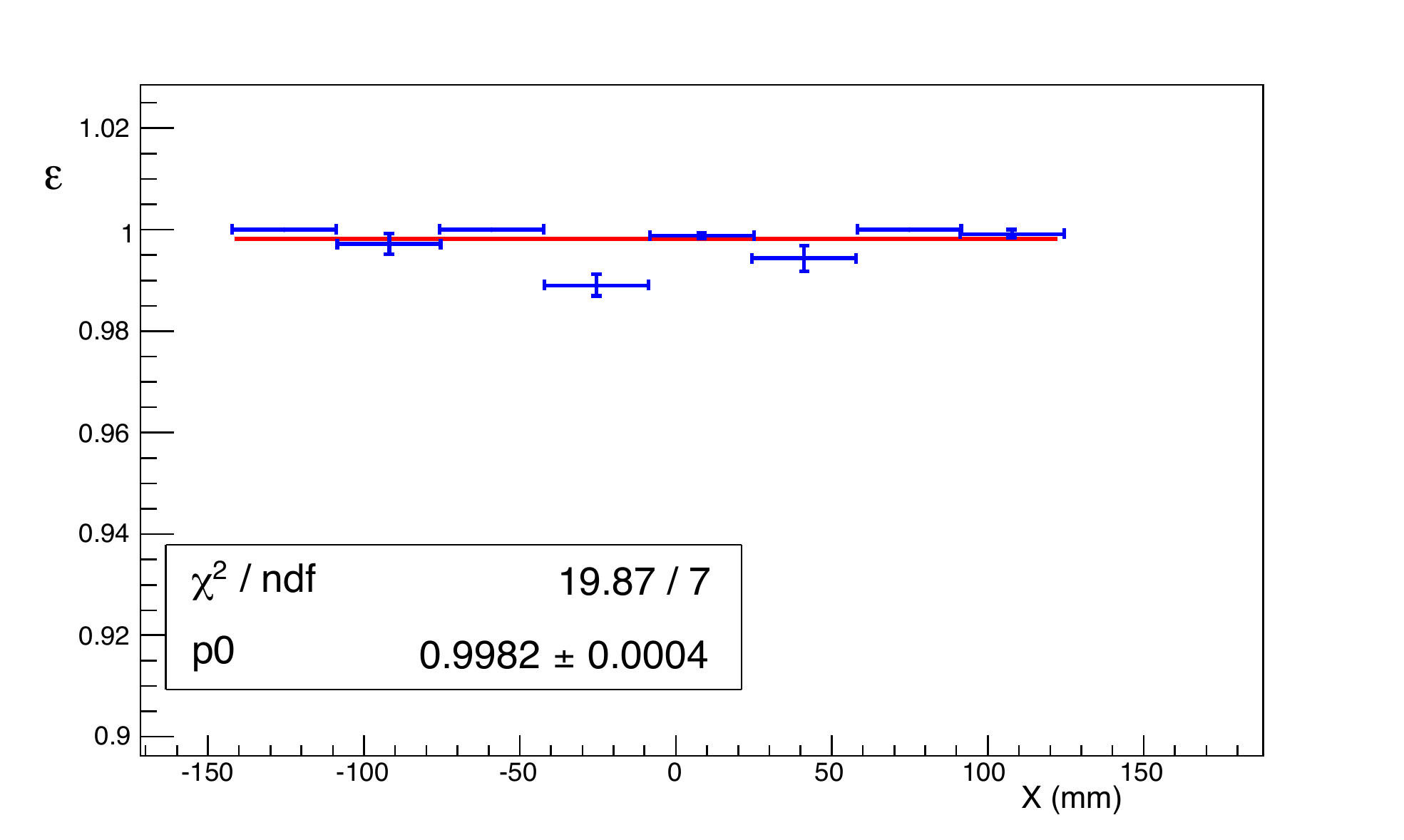}
\caption{Single layer efficiency measured on data.}
\label{effx_prima_staz}
\end{figure}

When the beam is present, CHANTI is hit by secondary particles produced
primarily by the interaction of the beam with the third station of the beam
spectrometer (GTK3) as well as by the penetrating particles of the beam halo, that
are essentially muons proceeding in a direction parallel to the beam and crossing the CHANTI stations perpendicularly to their plane frames. So
two types of events are recorded by CHANTI. Due to the triangular section of
the bars (figure \ref{bar} on the right), in the case of muons an anticorrelation
exists between the amplitudes, and then between the $ToT$, of the signals
recorded in two coupled bars. Since the $ToT$ has a logarithmic dependence
from the charge generated by the particle that passes through the bar,
a pseudo-charge $Q_{p}$ may be associated to each signal, defined as $Q_{p}=\exp \left( ToT/ToT_{0}\right)$, with $ToT_{0} \equiv 38$ ns.
This formula has been determined by the MC program that simulates in
detail the CHANTI. It can be used to determine the position where the
particle impacts two coupled bars doing a charge weighted average of the
nominal positions of the bars and, thus, to study the spatial resolution of
the detector. In figure  \ref{diff_2_5} it is shown, as an example, the difference
between the x-coordinate (vertical bars) determined by the positions of two
coupled bars of the third station and the corresponding one of the fourth
station. 
Assuming, as it is reasonable to do, the two stations have the same
spatial resolution, from that plot one obtains $\mathit{\sigma _{x}}=2.6$ mm.
A systematic study of all stations led to the result $\mathit{\sigma _{x/y}}\simeq
2.5$ mm. This has to be compared with the rms (4.9 mm) of a flat distribution with a pitch of half of a bar base.
Using tracks of passing particles ("muons") parallel to the beam direction,
the CHANTI efficiency was studied, for each station and for each view, both
for coordinates x (view with vertical bars) and y (view with horizontal
bars) separately and as a function of the xy position. Let us consider, for
example, the study of efficiency as a function of x. A "muon" is defined as
an event that gives at least one hit in two vertical coupled bars of a
station and in the corresponding bars of four other stations; the sixth
station is defined efficient if it presents for the same event at least one
hit in the corresponding coupled bars or in one of the two immediately
nearby bars. Efficiency is defined as the ratio between the number of times
the station was efficient and the total number of selected events. Figure \ref{effx_prima_staz}
shows the efficiency of the X layer of the first
station as a function of the position of the vertical bars. The single layer
efficiency, defined as above, resulted $>0.99$. 
It is worth noting that a layer efficiency $\geq 0.99$ ensures that the CHANTI inefficiency for detecting a charged particle, passing through both layers of a station, is less than 0.1\%. 

\section{Conclusions}
The  design, construction and test of CHANTI, a charged particle detector made of scintillation counters read by Silicon Photomultipliers, have been described. CHANTI will be used as a veto counter in the NA62 experiment at CERN.  Measurements done on prototypes and during the first period of the data taking with the complete detector resulted in a single channel resolution of 1.14 ns and a single layer efficiency greater than 0.99 in agreeement with the design values.

\acknowledgments
We want to thank the technicians of mechanical and electronic workshops of the Naples Section of the National Institute of Nuclear Physics (INFN) for the efforts they put in the development and construction of the detector. We want also thank A. Orlandi and A.Viticchie' from the National Laboratory of Frascati for their help in polishing and mirroring of the WLS fibers.
This work has been partially funded by Italian Ministry of University and Research (MIUR) with the PRIN Grant Prot. Number 2010Z5PKWZ\_008.

\end{document}